%% file: main.tex
\pdfoutput=1
\documentclass[preprint]{elsarticle}

\usepackage[margin=3cm]{geometry}

\usepackage{amssymb}
\usepackage{amsmath}
\usepackage{hyperref}
\usepackage{multirow}
\usepackage{rotating}

\usepackage[color=pink,textsize=small]{todonotes}

\DeclareUnicodeCharacter{2032}{'}
\DeclareUnicodeCharacter{02B9}{'}
\journal{TBC}
\begin{document}
\begin{frontmatter}

\title{Machine learning superalloy microchemistry and creep strength from physical descriptors}
\author[inst]{Patrick L. Taylor\corref{cor}}
\author[inst]{Gareth Conduit}
\affiliation[inst]{Cavendish Laboratory, University of Cambridge, CB3 0HE, UK}
\cortext[cor]{Corresponding author at: Cavendish Laboratory, University of Cambridge, CB3 0HE, UK. Email: pt409@cam.ac.uk}

\begin{abstract}
    We propose an element-agnostic set of descriptors to model superalloy properties with Gaussian process regression. Furthermore, we develop a correction method to deliver the best and most physical predictions for microchemistry in multi-phase alloys.
    The models' performance in predictions is confirmed for superalloy microchemistry, microstructure, and strength properties. When including new, unseen elements in the test data, the models still give good predictions; such extrapolations into new chemical-space would be impossible with component-based descriptors. 
\end{abstract}


\end{frontmatter}

\section{Introduction}
\input{sections/introduction}

\section{Computational method}
\input{sections/comp_method}

\section{Results}
\input{sections/results}

\section{Conclusion}
\input{sections/conclusions}

\section*{CRediT authorship contribution statement}
\textbf{Patrick Taylor}: Conceptualisation, Methodology, Software, Formal analysis, Data curation, Writing---original draft. \textbf{Gareth Conduit}: Conceptualisation, Methodology, Writing---review \& editing, Supervision.

\section*{Declaration of competing interest}
Gareth Conduit is a Director of materials machine learning company Intellegens. The authors have no other financial interests or personal relationships that could have appeared to influence the work reported in this paper.

\section*{Acknowledgements}
Patrick Taylor acknowledges the financial support of EPSRC and an ICASE award from Dassault Syst\`emes UK. Gareth Conduit acknowledges the financial support of the Royal Society. We would also like to thank Dr.\ Victor Milman and Dr.\ Alexander Perlov at Dassault Syst\`emes UK for their feedback on the manuscript. 
There is Open Access to this paper and data available at https://www.openaccess.cam.ac.uk/.

\bibliographystyle{elsarticle-num}
\bibliography{references}

\end{document}

%% file: sections/introduction.tex
Hume-Rothery first developed his eponymous set of rules in 1935~\cite{Hume-Rothery1935OnAlloys,Hume-Rothery1955AtomicMetallurgy.}. They describe whether any two elements could alloy together to form a stable solid solution. For substitutional alloys there are four rules: they concern the similarity of atomic radius, crystal structure, valency, and electronegativity. 
Further developments to solubility rules include the Pettifor scale and its 2016 proposed update by Glawe et al~\cite{Pettifor1984AMaps,Glawe2016TheMining}. 
The Hume-Rothery rules highlight the opportunity for fully element-agnostic models of phase composition that would enable them to be applied to any material, even those containing elements not before considered. 
However, the Calculation of Phase Diagram (CALPHAD) methodology---which has become the industry standard approach due to the promulgation of software such as ThermoCalc---typically relies on thermodynamic models that have been constructed element-by-element, limiting their broad applicability~\cite{Kattner2016TheDevelopment,Andersson2002Thermo-CalcScience,Bajaj2011TheMethod}. 

The same duality exists in applications of machine learning to alloys. 
Some researchers have adopted the approach of using alloy components directly as descriptors in their ML models. 
Such models have been used to model Ni-based superalloy microstructures, and design high entropy alloys and superalloys for a diverse range of applications~\cite{Harada1999DesignSuperalloys,Taylor2022MachineMicrostructure,Conduit2017DesignNetwork,Conduit2019ProbabilisticDeposition}. 
Other researchers have taken the approach of mapping alloy components to physical descriptors based on domain knowledge~\cite{Xue2017AnAlloys,Mamun2021AAlloys,Lee2022PhaseLearning}. 
This strategy greatly improves models trained on small datasets, whereas models trained on suitably large datasets were already able to self-encode the physical descriptions of a system in their latent space~\cite{Murdock2020IsProperties}, see Hart et al (2021) for a thorough review~\cite{Hart2021MachineAlloys}.

Superalloys are precipitation hardened---their important bulk properties, such as high-temperature creep and yield strength, derive from their two-phase microstructure~\cite{Durand-Charre1997TheSuperalloys,Reed2006TheApplications,Crudden2014ModellingSuperalloys,Fleischmann2015QuantitativeSuperalloys,Dodaran2020EffectSuperalloys}. Hence, accurate prediction of the relative phase fraction and phase compositions is a crucial first-step towards property prediction. 
However, each generation of superalloys has typically included additional elements~\cite{Sims1984AMetallurgists}, the most recent being the inclusion of ruthenium~\cite{Tin2004AtomicSuperalloys,Reed2004IdentificationTomography,Hobbs2008TheSuperalloys,Tsuno2009EffectMPa,Wang2014TheTemperatures}. 
The trend of additional elements improving properties is naturally extended by high-entropy superalloys (HESA)~\cite{Tsao2015DevelopingSuperalloys,Chen2018AProperties,Joele2021AHESA,Detrois2019DesignMatrix,Manzoni2019NewApproach}. 
Machine learning models that better transfer their inherent knowledge of alloy chemistry to new composition-space could accelerate the further development of superalloys. 

We propose a physics-inspired set of descriptors to describe superalloys that is element-agnostic. Gaussian process regression machine learning is then used to predict superalloy properties. For phase composition we build on previous work~\cite{Taylor2022MachineMicrostructure} to develop and demonstrate a probabilistic correction method. In addition to interpolative scoring of the models in the usual cross-validated manner (via a withheld, randomly selected, test dataset), models are also scored on their ability to extrapolate to datasets containing alloy components not seen during training. 
Finally, creep strength models of Ni superalloys are developed using similar physics-inspired descriptor sets. 

%% file: sections/comp_method.tex
In this section we detail the computational method developed to predict phase behaviour. 
We first describe the physical descriptor-set used as input features to our machine learning models, then proceed to give details of the Gaussian Process Regression models. 
Next, we discuss some particular problems with representing phase microchemistry data in machine learning models, and the novel correction method developed to address them. 
Finally, we describe our approach to physical descriptor GPR models of creep strength. 

\subsection{Data representation}
\label{subsect:data_rep}
Our first task was to predict superalloy microchemistry. This is described by the phase fraction, $f^\phi$, and the corresponding composition of each phase, $x_i^\phi$. Following other authors and our previous work, we adopt partitioning coefficients in place of phase composition, $k_i^\phi=x_i^\phi/x_i$~\cite{Taylor2022MachineMicrostructure,Harada1999DesignSuperalloys,Harada1988PhaseSuperalloys,Ofori2004ASuperalloys,Tin2004AtomicSuperalloys,Fuchs2002ModelingSuperalloy,Ma2007DevelopmentSuperalloys,Wang2014TheTemperatures}. The partitioning coefficients give a simpler encoding of physical information---for a given element, its partitioning coefficients across many different alloys are typically more similar than its various phase component percentages. In order to use variables that are more normally distributed, we use a further transformation of $f^\phi$, $k_i^\phi$ from their respective intervals to the real line:
\begin{align*}
    p_i^\phi &= \ln\left(x_i^\phi/x_i\right) \quad, \\
    q^\phi &= \mathrm{arctanh}\left(2f^\phi+1\right) \quad. 
\end{align*}
As well as being necessary mappings, they also capture the physical symmetry: $f^\phi\rightarrow1-f^\phi\iff q^\phi\rightarrow-q^\phi$ and $k^\phi\rightarrow {k^\phi}^{-1}\iff q^\phi\rightarrow-q^\phi$.

\begin{figure}[!ht]
    \centering
    \includegraphics[width=0.9\columnwidth]{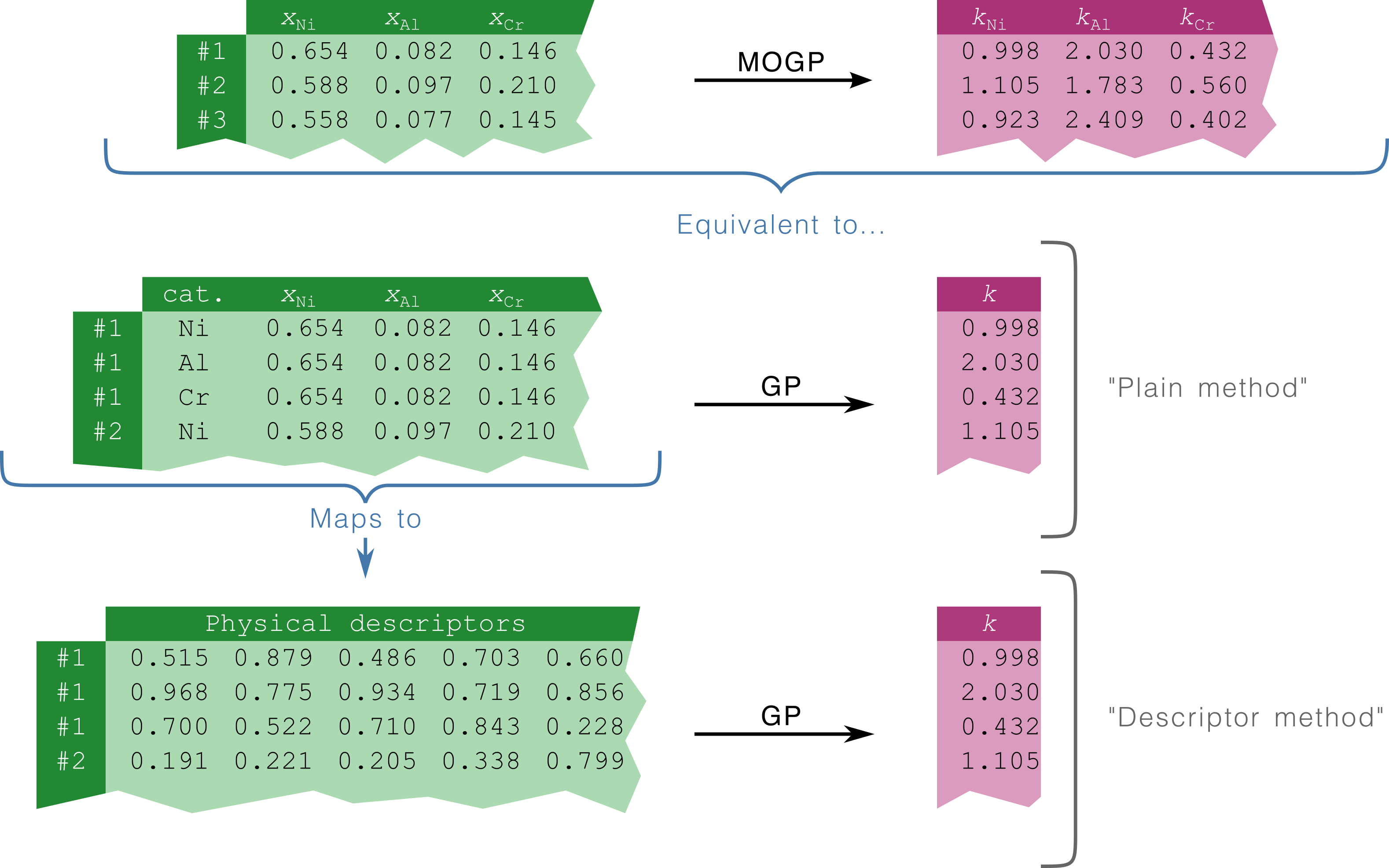}
    \caption{An overview of how the Gaussian process regression models in this work were formulated. }
    \label{fig:gpr_models_diagram}
\end{figure}

The dataset used in this work is an updated version of that in Ref.~\cite{Taylor2022MachineMicrostructure}, containing 123 entries with complete microstructure data (matrix and precipitate phase composition and fractions)
~\cite{Bagot2017AnSuperalloy,Basak2017MicrostructureSLE,Blavette1986AnSuperalloys,Blavette1996Atomic-scaleSuperalloys,Collier1986EffectsProperties.,Diologent2004OnSuperalloys,Dreshfield1970EstimationDiagram,Durand-Charre1997TheSuperalloys,Duval1994PhaseMC2,Glas1996OrderSuperalloys,Goodfellow2018GammaSuperalloy,Harada1988PhaseSuperalloys,Jalilvand2013InfluenceIN-738LC,Janowski1986TheMAR-M247,Khan1984TheSuperalloy,Kriege1969TheAlloys,Lapington2018CharacterizationSuperalloys,Llewelyn2017TheAlloys,Loomis1972TheSuperalloys,Ma2007DevelopmentSuperalloys,Miller1994APFIMSuperalloy,Park2022AlloyBehavior,Parsa2015AdvancedSuperalloys,Reed2004IdentificationTomography,Royer1998InSuperalloy,Schmidt1992Effect99,Segersall2015Thermal-mechanicalAlloying,Sudbrack2006TemporalAlloy,Sulzer2020TheApplications,Tin2004AtomicSuperalloys,Wanderka1995ChemicalMicroscopy,Wlodek1996TheDT,Yoon2007EffectsObservations,Zhang2000PhaseDeformation}. 
The phase composition dataset is presented in the conventional form, with the nominal alloy compositions and heat treatments being the inputs to our model, and the phase fractions and respective compositions being the output. However, in this work we ``reshape'' the dataset into a single-target format. This is represented graphically in Fig.~\ref{fig:gpr_models_diagram} as a two-step process. First the dataset is reshaped so that for a single-phase ML model, each partitioning coefficient is considered to be a separate output, and the input gains an additional column labelling the output element. Note that in the GPR framework, this is equivalent to a multi-output Gaussian process (MOGP)~\cite{Alvarez2011KernelsReview}. We will refer to this model as the ``plain method''. 
Secondly, the data is transformed into a physical representation, which is a function of composition and label, but does not explicitly preserve the label as a feature. This means that predictions can be made for labels (i.e. components) not present in the training data---so long as the physical descriptors are carefully chosen. We will refer to this model as the ``descriptor method''.

\begin{figure}[!h]
    \centering
    \includegraphics[width=0.9\columnwidth]{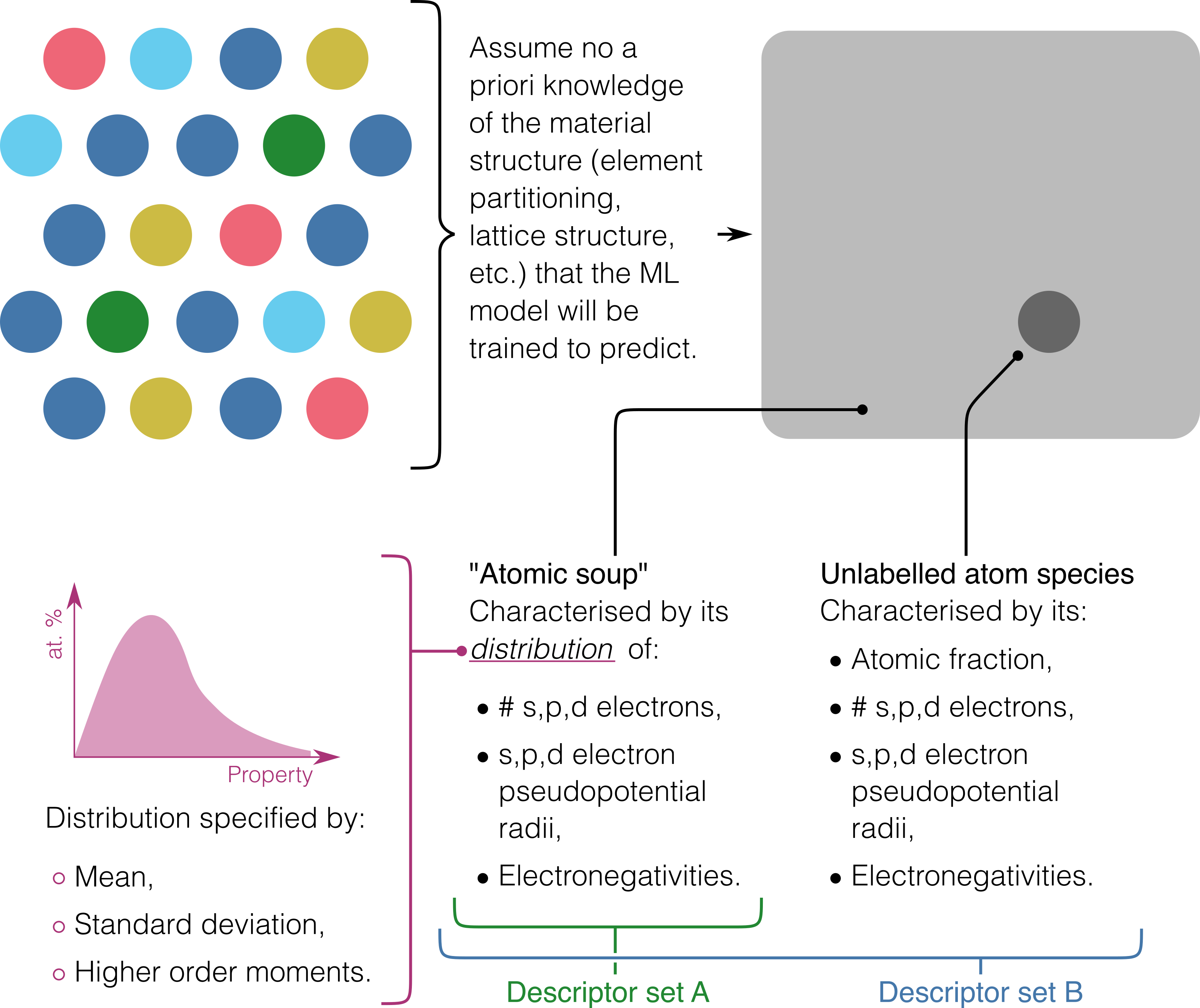}
    \vspace{5mm}
    \caption{How physics-based descriptors were selected for this work. Descriptor sets A and B were used for different aspects of microstructure modelling. Descriptor set A was used to model phase fraction, B was used to model the partitioning coefficients. }
    \label{fig:descriptor_diagram}
\end{figure}

\subsection{Physical descriptors}
\label{subsect:phys_features}
To fully understand the phase behaviour we must predict both the phase fraction, and also the partitioning of each element between the phases. The model for each property requires different inputs: 
for the phase fraction model, inputs were composition and heat treatment data; for the partitioning coefficient model, each entry gains a further feature corresponding to the component-label of the target, see Fig.~\ref{fig:gpr_models_diagram} and Section~\ref{subsect:data_rep}. In our model, both this label and the composition are converted into element-agnostic descriptors. 
The most natural choice of physical descriptors would be based on the lattice structure and atomic arrangement of an alloy, but these are a priori unknown. Instead, a given alloy---for which, in our case, we only know the nominal composition---can be thought of an ``atomic soup'', see Fig.~\ref{fig:descriptor_diagram}. Inspired by ab initio electronic structure methods used in physics and chemistry, the atomic soup can be represented by the distribution of the constituent atoms' electronic properties. Said distributions can be approximately specified by their mean, standard deviation; and higher order moments if necessary. This is how the descriptors (descriptor set A in Fig.~\ref{fig:descriptor_diagram}) for modelling phase fraction were formulated. For the GPR model of partitioning coefficients, the atomic species labels were also transformed to physical descriptors (Fig.~\ref{fig:gpr_models_diagram} and Section~\ref{subsect:data_rep}), chosen following a similar logic (descriptor set B in Fig.~\ref{fig:descriptor_diagram}). 

Our choice of descriptors bore a strong similarity to the popular descriptor-set MAGPIE~\cite{Ward2016AMaterials}. Those used by Ling et al~\cite{Ling2018MachineOptimization} and Liu et al~\cite{Liu2020PredictingLearning} to model nickel superalloy properties are also similar. 
Much like these descriptor sets, we found electronic structure inspired descriptors to be especially powerful. Data used to construct our descriptors was taken from Refs.~\cite{Waber2004OrbitalIons,2021Electronegativity}.

Alongside its nominal composition, the precipitation heat treatments applied to an alloy will also affect its final composition and properties~\cite{Taylor2022MachineMicrostructure}. Up to three heat treatment stages were used as input features for each alloy; each comprising a treatment temperature (${T_j}$) and time (${\tau_j}$), for a total of six features. 
The theory of Ostwald ripening gives a theoretical dependence on heat treatment time and temperature for an alloy's phase compositions~\cite{Lifshitz1961TheSolutions,Morral1994ParticleAlloys,Philippe2013OstwaldAlloys,Mostafaei2021ImprovementRatio}:
\begin{align*}
    \langle \Delta x_i^\phi \rangle^3(t) \sim \frac{\mathrm{e}^{-\varepsilon_i/k_\mathrm{B}T}}{T}t \quad ,
\end{align*}
where $\varepsilon_i$ is an activation energy associated with the diffusion of species $i$. Our model's targets are the log partitioning coefficients; heat treatment times in our database varied over several orders of magnitude (from around 15 mins to 1000 hours). Putting this altogether, we propose representing $m_\mathrm{HT}$-stage heat treatments with the following descriptors:
\begin{align}
    h'_j = 
    \begin{cases}
    \sum_{k=1}^j \ln(\tau_k) - \ln(T_k)\ , & j\leq m_\mathrm{HT} \\
    1/T_j, & m_\mathrm{HT} < j \leq 2m_\mathrm{HT} \quad .
    \end{cases}
    \label{eq:ht_transformation}
\end{align}
This retains the same total number of descriptors as in the initial input. In Section \ref{subsect:ht_results} we show that this set of heat treatment descriptors accurately capture the evolution of the alloys' microstructures.

\subsection{Gaussian process regression}
A large number of physical descriptors were proposed for use in our GPR model: much more than the number of input columns in our dataset (18 for our dataset with component and heat treatment inputs). 
To ensure that the GPR models were selecting the minimal set of relevant features during hyperparameter optimisation, an automatic relevance determination (ARD) Mat\'ern kernel was used~\cite{Duvenaud2014AutomaticProcesses}. 
In cases where the ARD kernel lengthscale $l$ for some of the features was found to be very large, $l\gtrsim10^3$, small improvements in the model's score could be obtained by retraining the model without using said features at all. 
The GPyTorch library was used for GPR~\cite{Gardner2018GPyTorch:Acceleration}. Both the L-BFGS and ADAM algorithms were tested for hyperparameter optimisation, with L-BFGS being found to give both better results for a small trade-off in training time. 

\subsection{Probabilistic correction to phase compositions}
\label{subsect:correction}
There are three physical constraints that alloy phase compositions---i.e. the output predictions of the GPR models for phase chemistry---must obey~\cite{Kattner2016TheDevelopment,Taylor2022MachineMicrostructure}:
\begin{align}
    \sum_i x_i^\phi &= 1 \quad, \label{eq:constraint_1} \\
    \sum_\phi f^\phi &= 1 \quad, \label{eq:constraint_2} \\
    \sum_\phi f^\phi x_i^\phi &= x_i \quad , \label{eq:constraint_3}
\end{align}
where there are $m$ phases labelled $\phi$ and $n$ components labelled $i$. Eqs.~(\ref{eq:constraint_1})\ \&\ (\ref{eq:constraint_2}) are hard constraints that the total concentration and total phase fractions must each sum to unity. 
Eq.~(\ref{eq:constraint_3}) can be interpreted as a soft constraint representing an assumption of minimal material loss in the forging process. Gaussian process regression does not present an obvious way to impose such constraints---in Ref.~~\cite{Taylor2022MachineMicrostructure} they were imposed on the final model outputs. In particular, a Bayesian approach was taken to apply constraint Eq.~(\ref{eq:constraint_3}) via a correction to the predicted phase fraction. In this work, the same approach is extended to apply a simultaneous correction to the phase fractions $f^\phi$ and the phase compositions $x_i^\phi$. The output of the GPR models for phase fraction and phase composition are taken to be independent Gaussian processes, which for each alloy gives a prior:
\begin{align}
    \exp\left[-\frac{1}{2}
    \left(\mathbf{q}-\hat{\mathbf{q}}\right)^T
    \mathbf{\Sigma}_q
    \left(\mathbf{q}-\hat{\mathbf{q}}\right)
    \right]
    \prod_\phi
    \exp\left[-\frac{1}{2}
    \left(\mathbf{p}^\phi-\hat{\mathbf{p}}^\phi\right)^T
    \mathbf{\Sigma}_{p^\phi}
    \left(\mathbf{p}^\phi-\hat{\mathbf{p}}^\phi\right)
    \right]
    \quad ,
\end{align}
and a likelihood relating to the soft constraint, where $\tau_i$ is a tolerance for `allowed' component loss, and $\sigma_i^{(3)}$ is an estimated uncertainty on the sum on the LHS of Eq.~\ref{eq:constraint_3}
\begin{align}
    \prod_i
    \exp\left[-\frac{1}{2}
    \left(\frac{
    x_i-\sum_\psi f^\psi x_i^\psi}
    {\tau_i x_i+\sigma^{(3)}_i}
    \right)^2
    \right]
    \ .
\end{align}
Combining these to give a posterior, expanding the exponent of the likelihood to quadratic order in $\Delta p_i^\phi = p_i^\phi-\hat{p}_i^\phi,\ \Delta q^\phi = q^\phi-\hat{q}^\phi$, and finally completing the square gives a new Gaussian probability distribution. 
This is equivalent to maximising the log-posterior probability with respect to the corrections to $\hat{\mathbf{q}}$ and $\hat{\mathbf{p}}$. This maximisation can be carried out subject to hard constraints for Eqs.~(\ref{eq:constraint_1})\ \&\ (\ref{eq:constraint_2}). Doing so produces a correction to each of $p_i^\phi$ and $q^\phi$, as well as a new, valid covariance for a given prediction, which in turn yields the associated uncertainties. 


\subsection{Creep strength modelling}
\label{subsect:creep_method}
We compiled a dataset of creep strength properties for single crystal Ni superalloys. Entries were drawn from academic literature and commercial databases
~\cite{Blavette1986AnSuperalloys,Caron2000HighApplications,Caron2008InfluenceSuperalloys,Diologent2004OnSuperalloys,Durand-Charre1997TheSuperalloys,Harada1993Atom-probeSuperalloy,Hopgood1986TheSuperalloy,Horst2020ExploringMicrostructure,Jovanovic1998MicrostructureSuperalloy,Khan1984TheSuperalloy,Khan1986EffectCMSX-2,Koizumi2004DevelopmentSuperalloys,Liu2020High-throughputResistance,Nathal1985Elevated100,Parsa2015AdvancedSuperalloys,Tian2012CreepTemperatures,Tian2012InfluenceRe,Tian2016InfluenceSuperalloy,Zhang2010EffectSuperalloy,1995High-temperature393}. 
A substantial contribution to the dataset was from the open source creep rupture life dataset compiled by Liu et al~\cite{Liu2020PredictingLearning} (266 entries). Database entries included multiple properties characterising creep strength, including elongation at rupture (82 entries), time to 1\% creep (79), and minimum secondary creep rate (55). There were significantly more entries available for creep rupture life (388), reflecting the importance of this particular property as the primary metric of creep strength~\cite{Xia2020MicrostructuralReview,Reed2006TheApplications}. For this reason we solely focused on modelling creep rupture life. 
Each entry had two corresponding experimental conditions, a temperature and applied stress. Other authors have found that direct ML models of the creep rupture life are more effective than modelling the Larson-Miller parameter~\cite{Mamun2021AAlloys}. We adopted the same approach, which meant both experimental conditions were included as input features. 

Three approaches were taken to create a GPR model for creep strength:
\begin{itemize}
    \item Plain composition descriptors (directly analogous to the plain method described above). 
    \item Physical descriptor set derived from the input composition only (analogous to the descriptors used for modelling phase fraction).
    \item Physical and metallurgical descriptors derived both from the input composition and from the fitted GPR microstructure model. 
\end{itemize}
Atomistic microstructural properties, in combination with precipitate morphology, are known to determine the physical mechanisms by which creep occurs~\cite{Durand-Charre1997TheSuperalloys,Reed2006TheApplications}. 
This motivated the use of the third model. 
In this model the precipitate fraction predicted by the microstructure model was used directly as a descriptor. 
Other derived descriptors were used: the lattice misfit between the two phases was calculated using the Vegard coefficients~\cite{Zhang2005TheCreep,Reed2009Alloys-By-Design:Superalloys,Neumeier2011LatticeSuperalloys}. 
The matrix phase stacking fault energies were approximated using fcc and hcp formation energies for each element~\cite{Chandra2012InfluenceRe,Tian2016InfluenceSuperalloy,Ma2007DevelopmentSuperalloys,Reed2006TheApplications}. The formation energies were calculated via density functional theory using the PBSESOL functional~\cite{Perdew2007RestoringSurfaces}. 
The mean and standard deviation of melting points for elements in the precipitate phase were included as proxies for the $\gamma'$ solvus~\cite{Grosdidier1998PrecipitationSuperalloys,Caron2000HighApplications}. 
The mean interdiffusivity metric and the mean metal d-level metrics used in the alloys-by-design procedure were also used as descriptors~\cite{Yukawa1988HighConcept,Reed2009Alloys-By-Design:Superalloys,Reed2016Alloys-By-Design:Superalloys,Tang2021Alloys-by-design:Manufacturing,Campbell2002DevelopmentSuperalloys}. 
The necessary microstructure predictions were made using the GPR model described in the previous sections, trained on the full microstructure database excluding any entries overlapping with the creep rupture life dataset. 

This GPR model incorporated more explicit high-level domain knowledge than any of the other models, which was only possible because the physical mechanisms that govern creep deformation have been well studied by metallurgists. However, the descriptors of this type that were used are not an exhaustive list, and for this reason physics-based descriptors that did not use explicit metallurgical domain knowledge were also used in this model (see Fig.~\ref{fig:crl_model_lengthscales}). 

%% file: sections/results.tex
With the data curated, descriptors selected, and machine learning formalism in place, we are now well-positioned to test the performance of our machine learning algorithms. We first study the performance of microstructure prediction, before studying the creep strength. 

\begin{table}[h]
\centering
\caption{Comparison of the root mean squared error (RMSE) for different components between the GPR model using descriptors and that using plain composition.}
\label{tab:result_summary}
\begin{tabular}{llccccccc}
 &  & \multirow{2}{*}{\begin{tabular}[c]{@{}c@{}}Phase frac. \\ (at. \%)\end{tabular}} & \multicolumn{6}{c}{Composition (at. \%)} \\
 &  &  & Ni & Cr & Co & Al & Ti & Heavy els. \\ \hline
Phase 1 & Descriptor method & - & 4.8 & 4.8 & 3.4 & 1.7 & 1.2 & 0.6 \\
 & Plain method & - & 5.4 & 5.2 & 3.2 & 1.6 & 1.3 & 0.6 \\
Phase 2 & Descriptor method & 4.4 & 3.8 & 2.7 & 2.7 & 2.4 & 1.5 & 0.5 \\
 & Plain method & 6.3 & 3.8 & 3.0 & 3.0 & 2.4 & 2.0 & 0.6 \\ \hline
\end{tabular}
\end{table}

\begin{figure}[h!]
    \centering
    \includegraphics[width=\columnwidth]{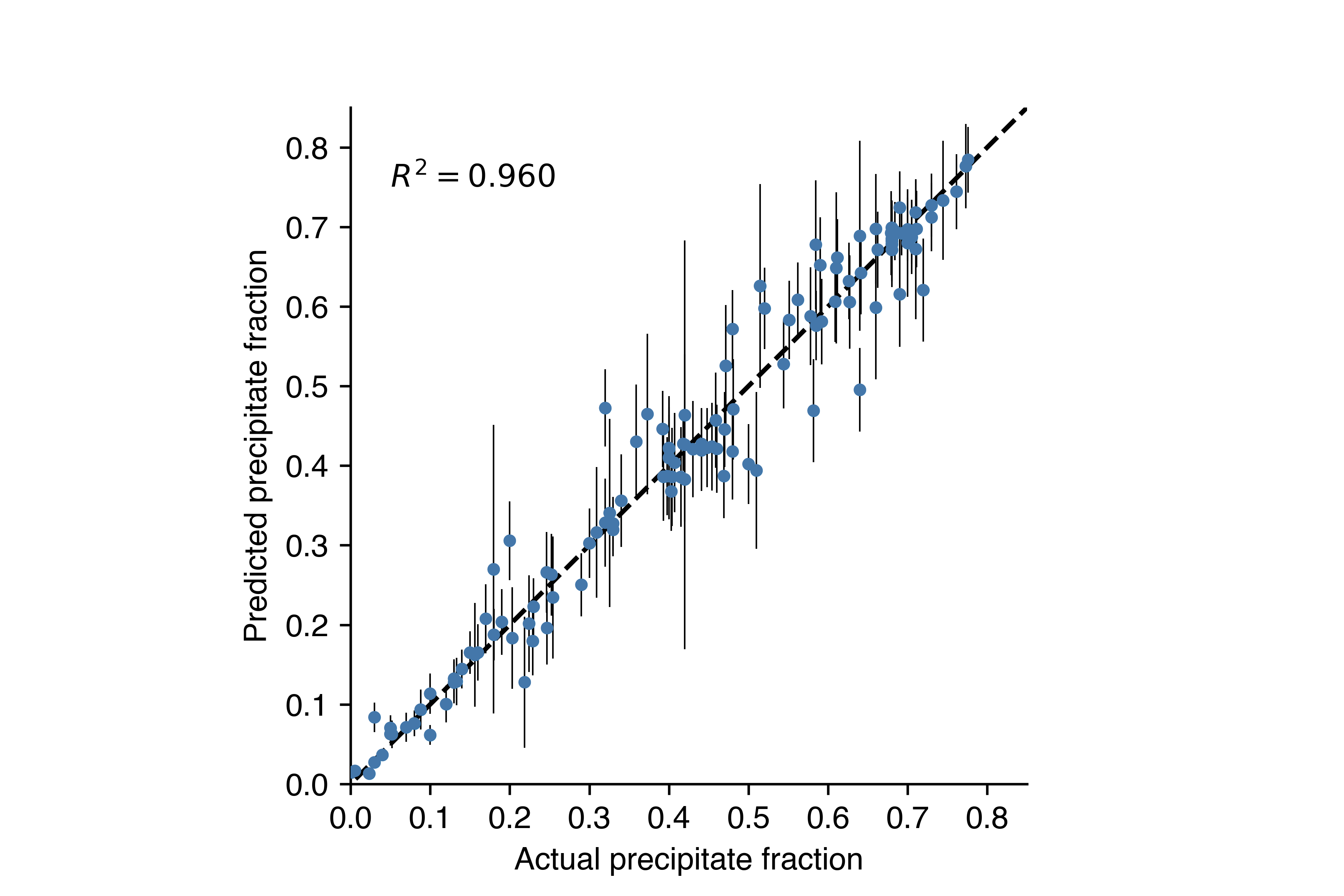}
    \caption{
    Predicted precipitate fractions from the descriptor model versus the actual values. The vertical bars are the model's uncertainties.}
    \label{fig:prc_frac}
\end{figure}

\begin{figure}[h!]
    \centering
    \includegraphics[width=0.9\columnwidth]{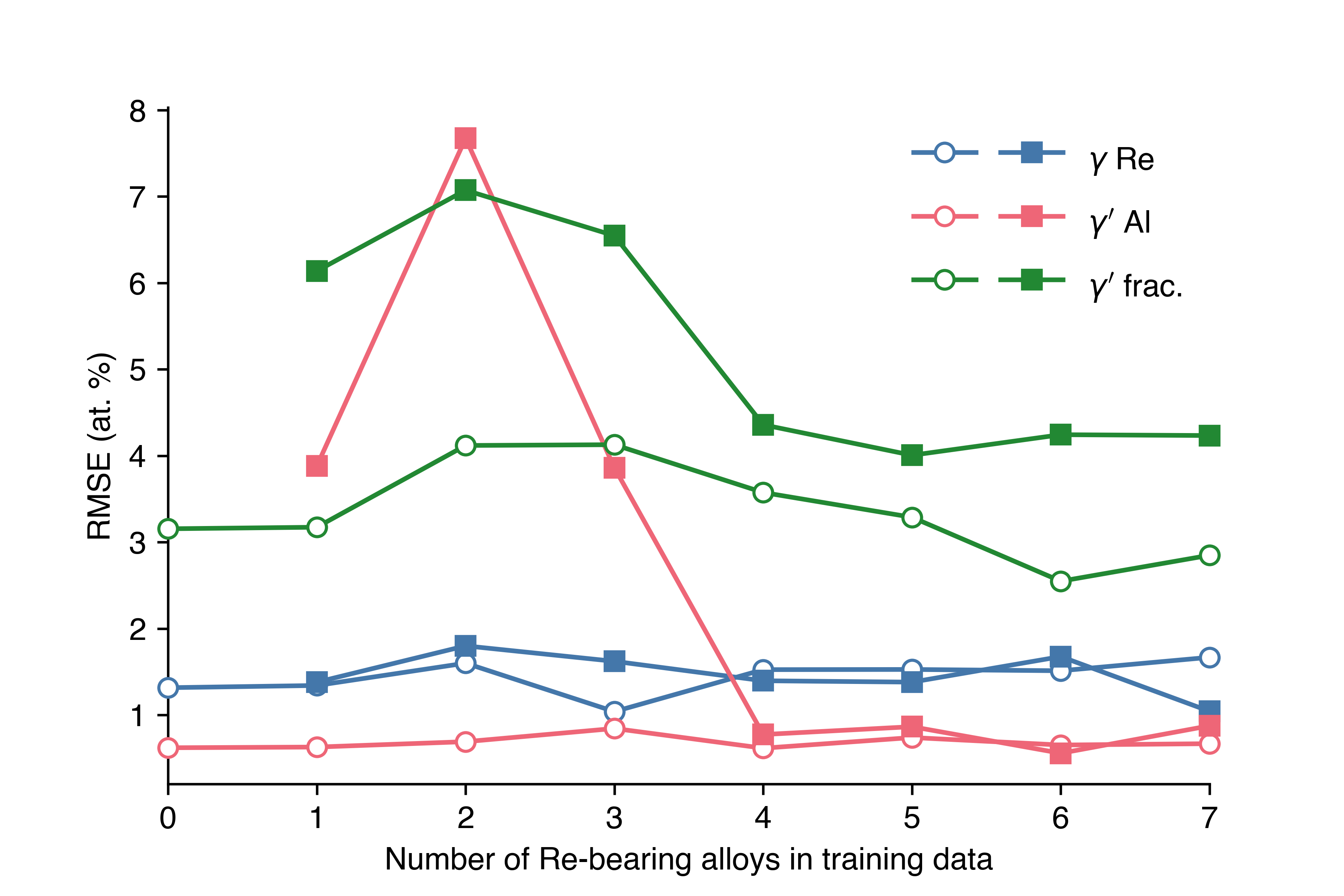}
    \caption{Predictions on a test dataset of Re bearing superalloys made by models trained on a dataset of 85 superalloys without Re plus a number of alloys with Re (indicated on the x-axis). 
    The points with open-circles are the physical descriptor model and with squares are the plain composition descriptor model.}
    \label{fig:ReRu_test}
\end{figure}

\begin{figure}[h!]
    \centering
    \includegraphics[width=\columnwidth,trim={0 15mm 0 0},clip]{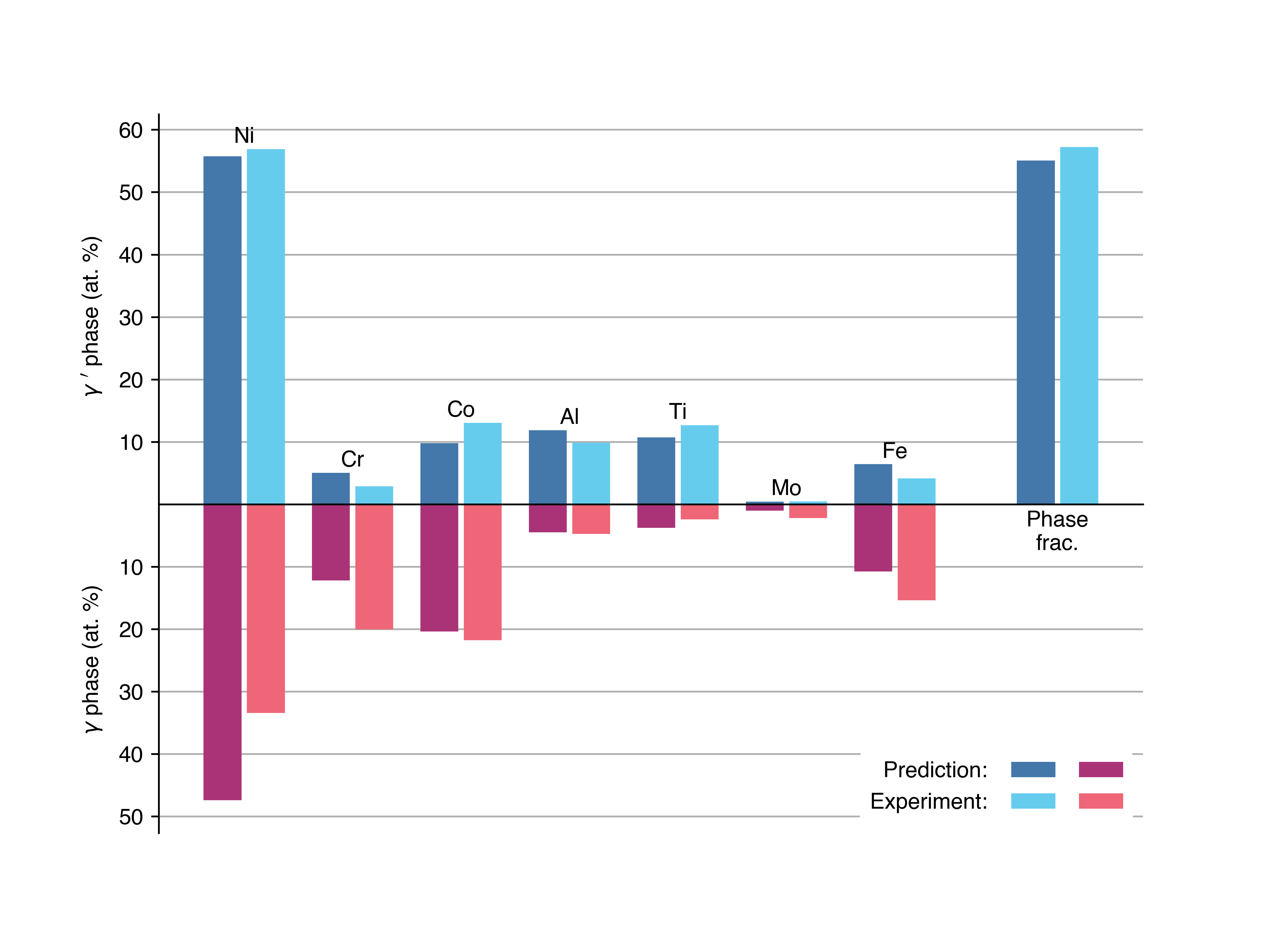}
    \caption{Predictions for the phase composition of a high-entropy superalloy (HESA)~\cite{Zhang2018Precipitation-hardenedProperties}  from the physical descriptor model.}
    \label{fig:hesa_composition}
\end{figure}



\begin{figure}[h!]
    \centering
    \includegraphics[width=\columnwidth]{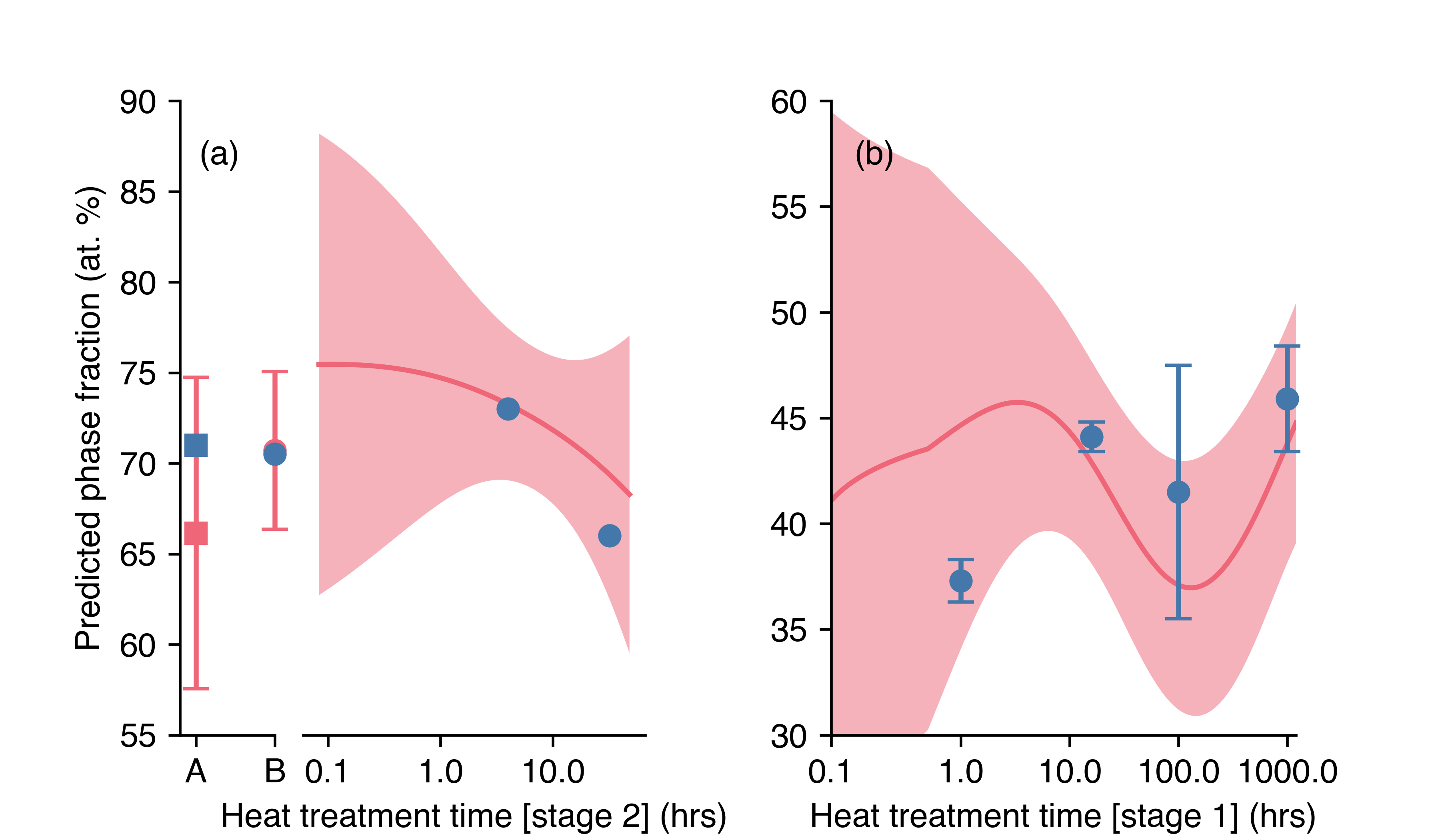}
    \caption{Predicted (red points/series) and experimental (blue points) data for phase fraction evolution with applied heat treatments in two superalloys. The vertical bars and shaded region give the standard uncertainty in the model's predictions. (a) the commercial superalloy SRR99. Point A is for a sample that has only had a solid solution heat treatment applied, point B has undergone a single precipitation heat treatment at $870^\circ\mathrm{C}$, and all following data points have undergone the same $870^\circ\mathrm{C}$ heat treatment preceded by a $1080^\circ\mathrm{C}$ heat treatment for the time indicated~\cite{Schmidt1992Effect99}. (b) an experimental superalloy.  Each alloy has undergone a single precipitation heat treatment at $760^\circ\mathrm{C}$ for the number of hours indicated~\cite{Goodfellow2018TheSuperalloys}.}
    \label{fig:heat_treatments_a-b}
\end{figure}

To compare the performance of the proposed physical-descriptor model to the plain model that uses composition features, we performed two rounds of tests on the microstructure data: firstly on blind validation data where all elements had been present in the training data, and secondly for extrapolating to new materials that contain fresh elements not present in the training dataset.

\subsection{Performance when all elements available}
\label{subsect:basic_microstruc_results}
We first test the performance of the physical descriptor model when information about all elements is available at training. Two models were trained: one that uses physical descriptors as inputs and a second that inputs the plain composition. 
We carried out ten-fold cross validation on the available microstructure data (123 database entries). For each fold, the model was trained via log-likelihood maximisation over the training dataset, then predictions were made on the withheld validation set. These results were combined to give a single set of predictions for the entire dataset. The correction method from section~\ref{subsect:correction} was applied in the same way to both models. 

The root mean squared error (RMSE) was used to compare the two models due to its ease of interpretation for percentage-like properties. The RMSE for each element and fraction component of the microstructure is given in Table~\ref{tab:result_summary}. 
The physical descriptor model was significantly better than the plain composition model for predicting phase fraction, improving the RMSE from 6.3\% to 4.4\%. The predicted versus actual phase fraction is plotted in Fig.~\ref{fig:prc_frac}, which confirms not only the quality of predictions, but also the accuracy of the uncertainty estimates. 
The phase fraction is the most crucial measure of a superalloy's microstructure given its physical influence on yield and creep strength. 
The descriptor-based model shows a smaller improvement for phase composition predictions when compared with the plain method, although it does still achieve the same or better RMSE for almost every element. 

We also confirmed the selected kernel hyperparameters. A Mat\'ern kernel with smoothness parameter $\nu = 2.5$ gave the best results for the log partitioning coefficient models and the phase fraction model. The fact that a smoother kernel gives better predictions when using physical descriptors is because the the models are `forced' to find the best physical descriptors rather than relying on the kernel's complexity to fit the data. In turn, the combination of a simpler, smoother model with a superior `understanding' of chemistry allows the model to extrapolate well.

\subsection{Extrapolative predictions in composition-space}
To test whether descriptors improve the ability of the model to extrapolate in composition-space, we adopted the simple approach of testing our GPR model on superalloy families outside the training dataset. We focus on two categories of alloys: Re/Ru-bearing Ni-superalloys, and high-entropy superalloys (HESA). 

The addition of Re and Ru to commercial superalloy compositions was one of the key innovations of the most recent generations of conventionally developed single-crystal superalloys~\cite{Sato2006TheSuperalloys,Argence2000MC-NG:Vanes,Walston2004JointSuperalloy,Reed2006TheApplications}. 
Here we repeat the training process and start from a database comprising 88 historic superalloys that contain neither Re nor Ru. Of the data on contemporary superalloys containing Re and Ru, 6 were randomly selected to form a test set, and the remaining were added incrementally to the training set so as to expose the importance of adding fresh alloys during a research project. 
Fig.~\ref{fig:ReRu_test} shows the RMSE vs.\ number of Re-bearing alloys in the training data. The initial predictions by the physical descriptor model are extremely impressive, giving about the same RMSE on the test data as it attains with any amount of Re-bearing alloy data. When a few data-points for Re alloys are added, the plain-descriptor model overfits so gives a higher RMSE than the physical descriptor model. For the crucial prediction of the $\gamma'$ phase fraction, this improvement persists up to the maximum number of additional training data points. 

High-entropy superalloys are a more recent development in alloy design, combining the design principles of high-entropy alloys and precipitation-strengthened alloys~\cite{Zhang2018Precipitation-hardenedProperties,Zheng2020AAlloy}. HESAs typically contain Fe as one of their entropy of mixing-boosting components. For a review on HESAs see Ref.~\cite{Joele2021AHESA}. 
HESAs are a prime target for the physical descriptor model owing to the large number of element permutations that cannot all be represented in the training dataset. 
The GPR model was trained on the full conventional superalloy database, that did not contain a single HESA entry, nor any entries containing Fe. The model was then tested on the HESA data collected by Zhang et al~\cite{Zhang2018Precipitation-hardenedProperties} that includes Fe-bearing superalloys, with results shown in Fig.~\ref{fig:hesa_composition}. The results are overall in excellent agreement, capturing the behaviour of the elements known to the model well, and notably making good predictions for Fe, which would have been impossible with the plain descriptor model. For both experimental HESAs, the RMSE for $\gamma'$ phase elements was 1.8\% and for $\gamma$ phase elements was 7.0\%. Once again, predictions for the $\gamma'$ phase are better than for the $\gamma$ phase, reflecting the stronger influence of physical factors on this phase's formation. Such a prediction would be simply impossible to make with a plain composition descriptor model.

\subsection{Heat treatments}
\label{subsect:ht_results}
To assess how well our heat treatment descriptors (Eq.~\ref{eq:ht_transformation}) capture the evolution of microstructure during a heat treatment, we test our model against two sets of experimental data~\cite{Schmidt1992Effect99,Goodfellow2018TheSuperalloys}. In each case, the physical descriptor model was retrained on the full database excluding the respective set of test alloys. 

The first test case was a commercial superalloy SRR99, aged under a variety of conditions. Our model captured both the qualitative and quantitative trends in the evolution of the $\gamma'$ fraction, see Fig.~\ref{fig:heat_treatments_a-b}(a). Surprisingly, it exhibited both the largest uncertainty and error for heat treatment A, the un-aged specimen. This could be reflective of a spurious correlation in the training data, since data for commercial superalloy microstructures are typically presented for fully heat treated specimens, whereas ``experimental'' alloy compositions are less frequently fully heat treated. 

The second test case was for an experimental five-component superalloy, where each specimen was aged at $760^\circ\mathrm{C}$ for increasingly long durations, shown in Fig.~\ref{fig:heat_treatments_a-b}(b). Our model captured not only the trend towards an increased $\gamma'$ phase fraction with treatment time, but also the observed decrease in precipitate fraction at intermediate ageing times. 
It captured this qualitative trend despite the fact that it is opposite to that observed for the alloy SRR99; which furthermore was one of the only alloys in the training data with more than two different heat treatments applied to the same composition. Our new proposed physical description of heat treatments, Eq.~\ref{eq:ht_transformation}, outperformed other descriptors for both test datasets, including a plain time and temperature descriptor and other Ostwald ripening-based descriptors, as well as giving better results for the five-fold cross-validation testing described above. 

\subsection{Creep rupture life model}
\begin{figure}[ht!]
    \centering
    \includegraphics[width=\columnwidth]{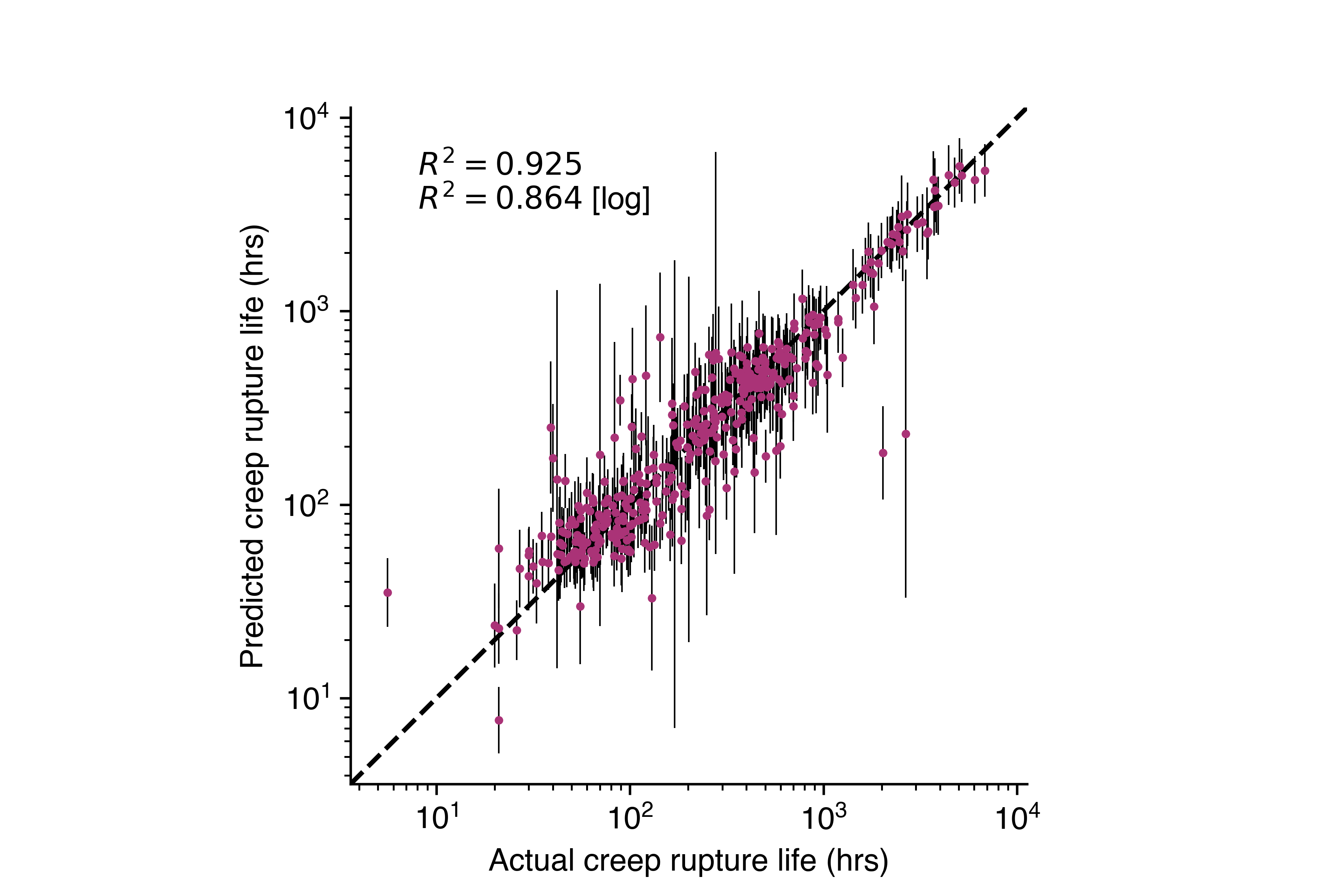}
    \caption{Predicted creep rupture life from a descriptor model including microstructure-based descriptors. The vertical bars are the model's uncertainties.}
    \label{fig:best_crl_model}
\end{figure}

\begin{figure}[ht!]
    \centering
    \includegraphics[width=0.8\columnwidth]{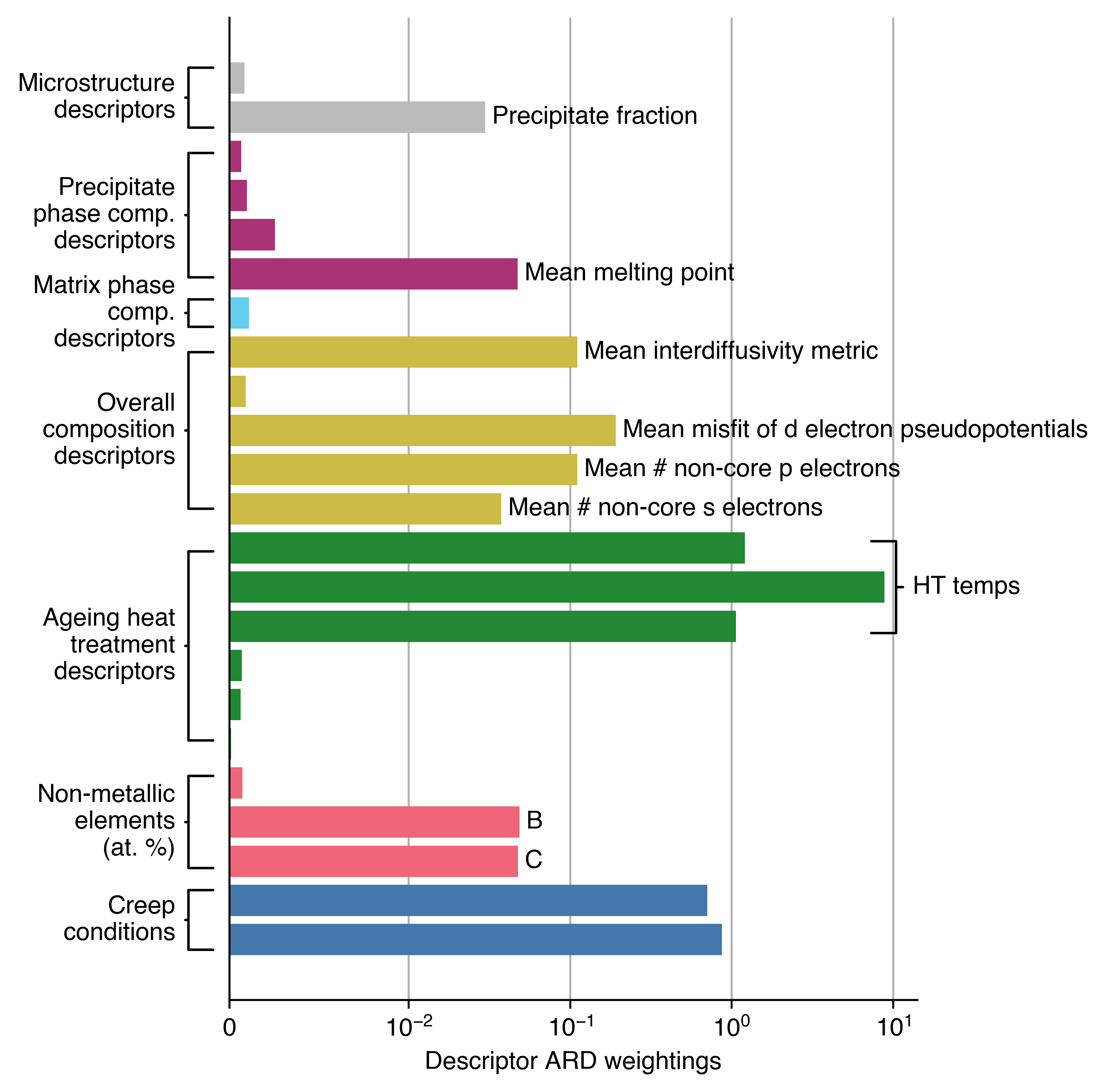}
    \caption{Feature weightings (inverse of ARD lengthscales) determined for a creep rupture life model trained on the full dataset.}
    \label{fig:crl_model_lengthscales}
\end{figure}

GPR models for the creep rupture life of signal crystal superalloys were trained. The accuracy of each model was assessed with ten-fold cross validation in the manner described in subsection~\ref{subsect:basic_microstruc_results}. As the creep rupture life data spans multiple orders of magnitude, $R^2$ scores are given for both the actual values and log values. 
An ARD Mat\'ern kernel was used in the Gaussian processes; but for the creep models, unlike those for microstructure, the optimal kernel smoothness parameter was found to be $\nu=1.5$. The decreased smoothness of the kernel is likely because creep rupture life data spans multiple scales and testing regimes. For example, it is known that creep in single crystal superalloys occurs by two main mechanisms, dislocation and diffusion creep, with the former mechanism dominating at low temperatures and the latter at high temperatures. 

\begin{table}[h!]
\centering
\caption{A summary of the coefficients of determination for GPR models of creep rupture life using different descriptors.}
\label{tab:crl_model_summary}
\begin{tabular}{lc}
Descriptors & $R^2$ \\ \hline
Physics + metallurgy-based descriptor method & 0.864 \\
Physics-based descriptor method & 0.856 \\
Plain method & 0.840 \\ \hline
\end{tabular}
\end{table}

Firstly, a plain composition descriptor GPR model was trained, which achieved $R^2=0.840$. Next, a model using a similar physics-based descriptor set to that used for microstructure, i.e. using descriptors calculated solely from each alloy's nominal overall composition, was trained. This delivered an improvement on the plain descriptor method, achieving $R^2=0.856$. 

This physics-based descriptor set was then further refined to include more high-level domain knowledge, summarised in Fig.~\ref{fig:crl_model_lengthscales}. Metallurgy-based descriptors developed in Section~\ref{subsect:creep_method} were added, some of which were calculated using phase compositions and fractions predicted by the pre-trained microstructure model described in the preceding sections. This model achieved a further improvement of $R^2=0.864$ (Fig.~\ref{fig:best_crl_model}). 

The relevance of various descriptors in this model as determined by the ARD kernel lengthscales are given in Fig.~\ref{fig:crl_model_lengthscales}. 
Many of the expected important descriptors are found to be relevant: the precipitate fraction, the overall mean interdiffusivity, and the mean melting point of the precipitate phase are selected. However, other descriptors that were expected to be highly relevant are not selected, including the lattice misfit, mean metal d-level, and the matrix phase stacking fault energy (SFE). 
The mean metal d-level is a metric used to estimate susceptibility to TCP phase formation during creep: commercial single-crystal superalloys have long been designed to avoid this behaviour, which will limit the influence of this descriptor for a model fitted to a dataset of largely commercial superalloys. The matrix SFE descriptor is unlikely to be sufficiently accurate: our estimate used an average of the difference between zero-point formation energies. This neglects thermal effects and uses an approximation to the SFE in fcc alloys. The lattice misfit is likely found to be irrelevant for both of those reasons. 
Of the non-metallic trace elements, the B and C at.\ \% are relevant, whereas Y is not. The same heat treatment descriptors used for the microstructure modelling were again used, but the Ostwald ripening inspired descriptors that were key to capturing microstructure evolution with ageing are not found to be relevant (although it should be noted that they indirectly enter the model via the predicted precipitate fraction descriptor). 
Three physics-based descriptors calculated from the nominal composition are found to be relevant. They are three descriptors that are also selected by the microstructure GPR model, so they may be relevant due to their influence on microstructure morphology, which is not explicitly modelled. 

%% file: sections/conclusions.tex
In this work we have proposed a set of physical descriptors for composition and heat treatment for use in machine learning models of superalloy microstructure. 
A model using physical descriptors outperforms a model using plain composition descriptors when making \textit{interpolative} predictions, see Table~\ref{tab:result_summary}. Furthermore, when making \textit{extrapolative} predictions, the model significantly outperforms the plain descriptor model, notably not suffering from such a severe overfitting effect (Fig.~\ref{fig:ReRu_test}). Moreover, it can also make predictions for alloys containing elements that were not even present in its training dataset (Fig.~\ref{fig:hesa_composition}). Such predictions are completely impossible to make using a plain descriptor model since they are not \textit{element-agnostic}. This means our model can make useful predictions for cutting-edge superalloys, such as high-entropy superalloys or superalloys containing new heavy elements. 

As well as standing up on its own, our model has a number of advantages over traditional CALPHAD. Previously identified benefits include the ability to easily retrain the GPR model, incorporate non-equilibrium features such as heat treatments, and most importantly its inherent quantification of uncertainties~\cite{Taylor2022MachineMicrostructure}. 
The model presented in this work captures both qualitative and quantitative effects of ageing on microstructure, see Fig.~\ref{fig:heat_treatments_a-b}(a--b). 
A further benefit of our model is that it is possible to easily incorporate computational data, whilst still treating it as distinct from empirical data. This can be achieved by simply adding an extra feature to the inputs: a binary descriptor encoding the method by which an entry has been obtained. The process of fitting the Gaussian process will determine how relevant the computational composition data is to predicting `true' compositions. 

In addition to the microstructure model, we developed a GPR model for the creep rupture life of single crystal nickel superalloys. This model builds on that developed for microstructure in two ways. Firstly, it explicitly incorporates predictions of the microstructure model as descriptors, and finds them to be relevant to making predictions, providing a direct use-case for the usefulness of our GPR approach. Secondly, it expands on the domain knowledge paradigm by making use of descriptors that encapsulate metallurgical theories and principles. By incorporating such high-level domain knowledge, the ARD lengthscales of the fitted Gaussian process can be used to infer greater physical understanding about the properties it models. With a coefficient of determination of $R^2=0.864$ for the log creep rupture lifes, this model is itself useful for making predictions of this key strength property of single crystal superalloys. Like the microstructure model, its predictions include uncertainties, a crucial feature for its intended usage in alloy design~\cite{Martin2017Profile-QSARCompounds,Irwin2020PracticalData,Taylor2022MachineMicrostructure}. 

Currently our model only calculates partitioning into two pre-determined phases, a use-case we have identified as most critical to superalloy design. Other authors have constructed machine learning models to classify alloys by phase stability~\cite{Lee2022PhaseLearning}. This points towards development of a fully-fledged, generic, and probabilistic approach to the Calculation of Phase Diagrams problem. Beyond alloys, this approach could be extended to phase separation in other systems including polymer/polymer, polymer/filler, and aqueous two-phase mixtures~\cite{Zhu2016ModellingMethod,Li2009ThermodynamicSystems,Li2015ThermodynamicSystem,Salabat2010LiquidliquidSalts}.